# Agile Methods: Testing Challenges, Solutions & Tool Support

Attique Ur Rehman
Department of Computer and Software Engineering, CEME, National University of Sciences and Technology, Islamabad, Pakistan
aurehman.cse19ceme@ce.ceme.edu.pk

Ali Nawaz
Department of Computer and Software Engineering, CEME, National University of Sciences and Technology, Islamabad, Pakistan
anawaz.cse19ceme@ce.ceme.edu.pk

Muhammad Abbas
Department of Computer and Software Engineering, CEME, National University of Sciences and Technology, Islamabad, Pakistan
m.abbas@ceme.nust.edu.pk

## ABSTRACT

Agile development is conventional these days and with the passage of time software developers are rapidly moving from Waterfall to Agile development. Agile methods focus on delivering executable code quickly by increasing the responsiveness of software companies while decreasing development overhead and consider people as the strongest pillar of software development. As agile development overshadows Waterfall methodologies for software development, it comes up with some distinct challenges related to testing of such software. Our study is going to discuss the challenges this approach has stirred up. Some of the challenges are discussed in this paper with possible solutions and approaches used for resolving these challenges. Also, the tools in practice are mentioned to improve the efficiency of the process.

## 1. INTRODUCTION

The trends of industry practices have been changing and updating lately. More automation techniques have been used and more advanced methods have been practiced for coping with the environmental challenges. Similarly, a new method has been practiced in the industry which helps for rapid software development named as Agile Method. Agile has been lately replacing the traditional software development methods and it has been practiced in the industry yet also having a great scope for research. As agile includes fast development techniques like Extreme Programming, it is also cost effective for any organization. It is different from the traditional waterfall development process as development in agile methodology is done according to sprints. Due to its difference with the traditional development, it has major challenges in software testing. A typical testing lifecycle involves different phases of identifying errors and faults and then correction of those errors and faults leading to testing again. As the development in the agile methodology is rapid and we therefore opt this methodology for rapid development of software so taking out time for comprehensive testing is difficult. Research has shown that agile methodology does not provides you this time for testing. We this study will identify the major challenges which may arise during agile testing and they can relate to the management of testing or can relate to the implementation of testing. The challenges can come at the organizational level between teams or can they can occur for the specific product. We also propose the solution for those challenges in light of the tools.

In this paper we have done a research about the challenges that arise in software testing for agile methodology. For this purpose, research papers from IEEE were consulted. There were certain challenges that were shortlisted to be mentioned in this paper and that are mentioned in section 2. Section 3 consists of the solution of the challenges described in the previous section followed by the description of tools that are mentioned in section 4. In the end we have mentioned the conclusion along with the future work.

## 2. Agile Testing
In software development when we talk about the word "agile", we usually refer to more flexibility, more collaboration and probably more simplicity [1]. And when we talk about the "agile testing", it normally refers to testing of bugs and errors in the workflow of agile development. But a reasonable question arises. Is this testing that easy and typical to implement? The answer to this question is that there are certain challenges which might occur in the way to perform testing during agile development. Following can be the challenges that might occur during testing.

### 2.1 Challenges
The following are the challenges of testing during agile development methodology

- Infrastructure for testing
- Test documentation
- Insufficient Test Coverage
- Broken code after frequent builds
- Early defect detection
- Insufficient API testing
- Lacking focused testing

These points in details are described below;

### 2.2 Infrastructure for testing
As we know that agile means that we will work in a short span time relatively than the other methodologies [2]. Under certain circumstances it is very difficult to create an environment for test execution which requires some user-centric/ customer-centric configurations. These configurations are basically cloning of the customer systems and they have identical hardware, operating system, third party software, and database. This would also include the challenge of installing the version of software under test on the customer centric configurations [3].

### 2.3 Test Documentation
Test documentation is considered as one of the major challenges in agile development. As in agile, the requirements of the customer are continuously changing and we are more focused on delivering the product rather than focusing on the documentation. Rapid programming is used for this purpose. Among all this, to keep hold of the testing documents, test data, test cases and scenarios, is a challenge that is to be worked upon [4].

### 2.4 Insufficient Testing Coverage
As mentioned above, due to continuous change in requirements and product, this method lacks stability. Do due to this continuous change, maintaining test coverage is difficult. Another major reason for insufficient test coverage is due to change in code. Major changes in code makes it difficult for the coverage to be maintained [5].

### 2.5 Broken Code after Frequent Builds
Due to frequent change and daily compilation there is a chance that code break occurs. This code break requires testing at each built which is obviously very difficult as the resources are not sufficient every time. Resources are always constrained. So automated testing can be a helping tool for this purpose [6].

### 2.6 Early Defect Detection
Due to frequent change in requirements, defect detection in early phases is majorly an issue which is hard to deal with. Defect detection in early phases is consequently less damaging and costly rather than detection during the production. This would lead to high cost and more over all damages [7].

### 2.7 Insufficient API Testing
Software these days are designed with SOA (service-oriented architectures) that expose their APIs so that developers can use the API services and extend the solution. So, for those developers which can develop an API, it is easy to test, but since in agile we face rapid development, mostly APIs are used, and testing the APIs is not easy. In fact, it requires strong coding skills and ample time [8].

### 2.8 Lacking focused testing
From the discussion on the above it is clear that there is no proper separate phase of testing in agile methodology. The center of focus is rapid development where in the developers rapidly develop the product and the requirements are constantly changing in order to fulfill the customer's needs. In all of this process there is a lack of focused testing. No specific phase is given to perform testing activities which creates problem for doing testing [9].

### 3. Solutions:
Agile is ideal for projects that are need to deliver in short interval of time, high capability of changing in requirements, capabilities of people working on the projects and technologies being used [10]. To overcome challenge and be a successful agile development team, teams need to implement following three aspects in their development and test infrastructure:

### 3.1 Self-service Provisioning and On-Demand Scalability

Agile developers should be able to gauge environments up and down easily in order to create, re-create, deploy, copy, delete and change the development with flexibility or test environment on demand of stakeholder without concerning IT assistance [11].

### 3.2 Library of Virtual Datacenter Templates

Agile developers' teams should generate a consistency in versions of the different releases such as customer-specific variations, current release stack and prior release stacks as templates. After reporting a major issue Agile teams should be able to immediately create a new environment that matches the suitable release state among the one mentioned above to duplicate, resolve, test, and deploy the issue of new code [12].

### 3.3 Complex Bug Capture and Reproduction

Quality Assurance and other support teams should be able to capture and quickly recreate complex situations, so that relevant resources can be managed as development team resources and should work to identify main causes of the problems and develop a solution, similar to the QA team moves forward with additional testing [13].

### 4. Test documentation
In traditional work when developing larger portions of the system documentation has to be expansive. But in agile projects the test plan is short and usually consist of one or two pages only that is why we can implement it with a testing strategy [4]. An alternative way is to implement checklists and exploratory testing to overcome documentation problem.

### 4.1 Testing strategy

Use of test strategy is an alternative way of describing test descriptions. It describes how the testing of the system is usually implemented. All of the typical test document descriptions can be described in the test strategy. With the assistance of this procedure we can have situations which are constantly significant yet not explicit to this run. The test methodology is a living archive and evolve after some time even with the small changes in the development team. These are some ordinarily utilized records in flexible testing [14]:

- A test strategy that defines how the software system is usually tested.
- A testing plan for each sprint.
- Specifications of testing which contain test cases.
- Test Ideas for experimental testing and test logs in which the consequence is distinguished.
- Checklists of installation testing and regression testing.

### 4.2 Checklists and exploratory testing
In many cases, checklists work better. Do not make huge checklists but makes it simple by describing in 1-2 lines what should be tested. Make a checklist per area to be tested and final checklist will look like a group of one liner.in addition you can also create a supplementary checklist for overall system having general test scenarios.

## 5. Insufficient Test Coverage

Research in agile testing exposed some critical techniques that agile teams would need to implement in order to overwhelmed coverage scenario [5]. The most significant techniques recognized are [15]:

1) Describe and implement "just-enough" acceptance tests. This technique boosts the confidence of everyone involved in the project by showing that user sections are complete and functional at the end of each sprint.
2) Automate the acceptance tests to 100 percentage as possible. This technique overcomes ever-growing manual regression testing and we get more coverage.
3) Make acceptance tests automatic by using a "subcutaneous" test method with a xUnit test framework [17].
4) Run all acceptance tests in the regression test suite [18]. This approach provides rapid feedback to the team.
5) Perform unit tests of all new code segment during a sprint. This approach will help to raises coverage and "just enough" acceptance testing.
6) Implement multiple builds per day with all created unit tests [19]. This approach allows more common integration of developer code and provide quicker response of possible integration issues.

## 6. Broken code after frequent builds

In 2.2.3 solution to coverage problem is describes as frequent builds with unit tests. That can cause broken code problem to overcome this problem an automatic testing tool can be used to perform regression testing whenever a new product is being developed. There are lots of tools available for continuous integration such as Cruise Control, Hudson and SmartBear's Automated Build Studio. Use of such tools will ensure and detect the constancy and integrity of the product through automated tests.

## 7. Early Defect Detection

There are two ways to accomplish early defect detection one is by Implementing peer reviews and the other is use of static analysis tools to generate report of program structure and defects as early as possible.

### 7.1 Peer Code Review

Preforming peer code review [20] on user stories, manual and automated tests. These reviews can be done via normal/common review techniques or we can use a tool like SmartBear's QA Complete to perform the review online and communicate progress and changes during the online review.

### 7.2 Reviewing Automated Tests

The agile testing teams should strongly implement the static analysis tools. Static analysis tools generate report automatically of program source code without generally executing the code. As we have to perform multiple builds, on every build running static code analysis will help developers to prevent defects early in the phase that might not be discovered until production.

## 8. Insufficient API Testing

To prevent challenge of API tests, there are tools explained in section 2.3 that allow developers and testers to generate test cases of the API without having strong coding knowledge [8], so this is another way to ensure that the implemented services are fully tested. Tool support is discussed in coming sections.

## 9. Lacking focused testing

Solution to this challenge is also explained in all previous sections of solution i.e. focused testing can be achieved by using automated testing tools and multiple builds using those tools.

## 10. Software Tools Used:

As mentioned above there are many tools that can solve the problems occurring in agile development methodology. We have shortlisted six of these testing automations tools which can be serving the purpose. The following is the list of tools along with the description.

- N-Unit
- STAF (Software Testing Automation Framework)
- Smart Bear
- Cucumber
- FitNesse
- Robot Framework

### 10.1 N-Unit

N-Unit is a special testing framework that only supports .NET framework for testing purposes. It was initially ported from Junit that actually supports JAVA. Initially N-Unit was a console-based runner but with the advancement to version 3.0 it provided a GUI to its user.

N-Unit initially supported only unit testing but aimed to do integration and acceptance testing as well. With the version 3.0 it removes this limitation and also supports integration and acceptance testing along with the unit testing. It also provides generation of tests for various codes [15], [16].

### 10.2 STAF

STAF stands for software testing automation framework. It is an easily available framework which supports multiple languages. It provides multiple services like process supplication, resource management, logging and monitoring. STAF is rather different than other tools in a way that it supports its users. It supports the users by allowing them to make their own software testing solutions which are automated and which are pluggable [15].

STAF can be very helpful in solving basic problems like frequent product cycles, less time, reduced testing and many more. As mentioned in the above section 2.1, the challenges of agile testing are caused due to the problems as mentioned before which are easily solvable by STAF. The latest version of STAF is 3.4.24.1 which was released on January 7, 2016.

### 10.3 Smart Bear

Smart Bear is latest testing tool which provides help in automating certain QA activities. We can say that it is a smart testing management system. Smart Bear provides QAC i.e. QA Complete which helps in performing certain QA activities. It provides easy testing environment. QAC has many features like central test case management which is responsible for providing the supervision of test schedules, create, manage and record defects, trace test to user stories and generate report of test execution. It can also give full visibility into the testing process and helps in assurance of shipping the software. QAC fits perfectly into your existing processes and it also makes informed decisions. This tool saves time and increases productivity [21].

### 10.4 Cucumber

Cucumber is a latest software testing tool that works to enhance software testing automation. The focus of cucumber is on its customers. It helps the customers in testing the documents. Testing requirements is even easy now. Moreover, cucumber supports behavior driven development. The focus of cucumber is on acceptance testing. It is a very powerful tool that can help in solving challenges of testing in agile [18].

### 10.5 Fit Nesses

FitNesse is an acceptance testing automation tool that helps the users in automating the acceptance testing. It was brought later after J Unit and N Unit but it claims to have much better and quick results.

FitNessse can give us the feedbacks very early and very frequently. Secondly Fit Nesses performs deterministic testing which means that it would either run green or red. FitNesse exercise more paths through business logic. It focusses on the acceptance testing as mentioned earlier and aims to test the requirements as well. It focuses on "Building the Right Code" [22].

### 10.6 Robot Framework

Robot frame work is another tool for software testing automation which supports acceptance testing and acceptance test driven development (ATDD). The test data provided in a tabular form which makes it easy to use. We can enhance its capabilities by adding certain libraries which are implemented in python and java.

Robot Framework is introduced on GitHub where further documentation and source code is available. Robot Framework is independent of operating system. The framework is implemented on python and runs on JVM and .NET. It was released under Apache License 2.0 and it is an open source software [23].

### 11. CONCLUSION:

Our primary contribution in this paper is the challenges agile software testing facing and the respective solutions for each one of them. Solutions gathered through research shows that most of improvement to testing agile developed software can be done through different tools. Six tools are identified in the previous section. Most of them support continuous builds and smart documentations.

Usually agile development we do not focus on writing test cases. But our study proposed that a little of documenting tests is needed to achieve the acceptance criteria. So, we should focus on documenting only what is needed. Also, the continuous builds using tools will improve the testing in short time.

**AUTHORS' BACKGROUND**

| Your Name | Position | Email | Research Field | Personal website |
|---|---|---|---|---|
| Attique Ur Rehman | Master student | aurehman.cse19ceme@ce.ceme.edu.pk | Software Engineering, Software project management | |
| Ali Nawaz | Master student | anawaz.cse19ceme@ce.ceme.edu.pk | Software Engineering, Machine Learning, Data Mining | |
| Muhammad Abbas | Associate Professor | m.abbas@ceme.nust.edu.pk | ERP Systems, Project Management | |